\documentclass[%
reprint,
superscriptaddress,
amsmath,amssymb,
aps,
]{revtex4-2}

\usepackage{graphicx}
\usepackage{dcolumn}
\usepackage{bm}

\begin{document}

\title{Effect of spin-dependent tunneling in a MoSe$_2$/Cr$_2$Ge$_2$Te$_6$ van der Waals heterostructure on exciton and trion emission}
%
%
\author{Annika Bergmann-Iwe}
\affiliation{Institute of Physics, Rostock University, 18059 Rostock, Germany}
\author{Swarup Deb}
\affiliation{Institute of Physics, Rostock University, 18059 Rostock, Germany}
\affiliation{Saha Institute of Nuclear Physics,
Kolkata, India}
\author{Klaus Zollner}
\affiliation{Institute for Theoretical Physics,  Regensburg University, 93040 Regensburg, Germany}
\author{Veronika Schneidt}
\affiliation{Institute of Physics, Rostock University, 18059 Rostock, Germany}
\author{Mustafa Hemaid}
\affiliation{Institute of Physics, Rostock University, 18059 Rostock, Germany}

\author{Kenji Watanabe}
\affiliation{Research Center for Electronic and Optical Materials, NIMS, 1-1 Namiki, Tsukuba 305-0044, Japan}
\author{Takashi Taniguchi}
\affiliation{Research Center for Materials Nanoarchitectonics, NIMS, 1-1 Namiki, Tsukuba 305-0044, Japan}
\author{Rico Schwartz}
\affiliation{Institute of Physics, Rostock University, 18059 Rostock, Germany}
\author{Jaroslav Fabian}
\affiliation{Institute for Theoretical Physics,  Regensburg University, 93040 Regensburg, Germany}
\author{Tobias Korn}
\email{tobias.korn@uni-rostock.de}
\affiliation{Institute of Physics, Rostock University, 18059 Rostock, Germany}

\begin{abstract}
	We study van der Waals heterostructures consisting of monolayer MoSe$_2$ and few-layer Cr$_2$Ge$_2$Te$_6$ fully encapsulated in hexagonal Boron Nitride using low-temperature photoluminescence and polar magneto-optic Kerr effect measurements. Photoluminescence characterization reveals a partial quenching and a change of the exciton-trion emission ratio in the heterostructure as compared to the isolated MoSe$_2$ monolayer. Under circularly polarized excitation, we find that the exciton-trion emission ratio depends on the relative orientation of excitation helicity and Cr$_2$Ge$_2$Te$_6$ magnetization, even though the photoluminescence emission itself is unpolarized. This observation hints at an ultrafast, spin-dependent interlayer charge transfer that competes with exciton and trion formation and recombination.
\end{abstract}

\maketitle
\section{Introduction}
In recent years, two-dimensional (2D) crystals and van der Waals (vdW) heterostructures~\cite{Geim-vdW-2013} consisting of different 2D crystals have been one of the most active fields in solid-state research. Besides graphene, the semiconducting transition metal dichalcogenides (TMDCs) such as MoSe$_2$ have garnered a lot of research attention. This is due to their exciting electronic and optical properties. In the monolayer (ML) limit, they become direct-gap semiconductors~\cite{Mak-MoS2directGap-2010} with a peculiar band structure leading to spin-valley coupling~\cite{Xiao-SpinValleyTMDCs-2012}. 
These properties make them potentially interesting for spintronics~\cite{Sierra-Review-vdW-Spintronics2021} and valleytronics~\cite{Schaibley-valleytronics-review-2016}, where information is encoded in the spin or valley degree of freedom of carriers, instead of their charge. In ML TMDCs, the optical selection rules allow generation of a coupled spin-valley polarization of excitons using circularly polarized excitation, and the excitonic valley polarization degree can be read out directly in helicity-resolved photoluminescence (PL) measurements~\cite{Mak-valleyPolMoS2-2012,Zeng-valleyPumpingMoS2-2012}. Depending on the specific TMDC material, this mechanism for generating a valley polarization can be very effective and robust, even for highly nonresonant excitation. 

For the specific case of MoSe$_2$, however, the valley relaxation rate is extremely fast compared to that of exciton recombination, so that a significant circular polarization of the PL emission can only be observed for near-resonant excitation~\cite{Wang-valleytimePLMoSe2-2015,Baranowski-Rates-ValleyPol-2017,Tornatzky-resonanceValleyMoSe2-2018}. Despite the unfavourable relaxation rates, an excitonic valley polarization can be achieved by lifting the energy degeneracy of opposite valleys. A usual strategy has been to exploit external magnetic fields perpendicular to the plane of a TMDC ML to break time-reversal symmetry. This introduces a valley Zeeman splitting and leads to a preferential occupation of the energetically favorable valley, even for unpolarized and nonresonant excitation~\cite{MacNeill-ZeemanMoSe2-2015,Mitioglu-MagnetoPL-WSe-2015}. However, the effective g factors for TMDC monolayers correspond to a splitting of only about 0.2~meV per Tesla, so that magnetic fields of several Tesla are required to achieve a significant valley polarisation even at liquid helium temperature. 

As such field strengths are impractical for device applications, the use of magnetic proximity effects~\cite{Scharf-magneticProximity-Excitons-2017}, where the proximity of a magnetic material induces a magnetization through exchange interaction, has been explored in recent years. This has made it possible to tailor the spin-valley properties in TMDCs without the need for an external magnetic field~\cite{Zhong-CrI-WSe-HS2017}. Alternatively, spin injection from ferromagnetic materials was used to generate valley polarization in TMDCs~\cite{Ye-SpinInjectGaMnAs-2016}. A variety of bulk ferromagnetic materials ranging from metals like Nickel~\cite{Tornatzki-MoSe2-Nickel2021} to ferromagnetic semiconductors such as Ga(Mn)As~\cite{Ye-SpinInjectGaMnAs-2016} and EuS~\cite{Zhao-EuS-proximity-2017} were employed in experimental studies.
However, the presence of surface states, dangling bonds, interface reconstruction, etc. greatly complicates their integration into vdW heterostructures. 

For this, layered ferromagnetic materials~\cite{Mak-review2Dmagnets2019,Zhang-review2Dmagnets2022}, made of chemically stable atomically thin sheets, have recently emerged as a viable alternative for integration into vdW heterostructures. While there have been a number of experimental studies~\cite{Zhong-CrI-WSe-HS2017,Lyons-MoSeCrBr-HS2020,Rahman_CGT-WS2021,Zhang-MoSeCGT-gated2022}, among the plethora of available TMDC and vdW ferromagnets, it remains elusive which material combinations, thicknesses, and other parameters offer the best material platform for studying phenomena such as spin-dependent tunneling and magnetic proximity effects.

Here, we report on an optical spectroscopy study of vdW heterostructures consisting of monolayer MoSe$_2$ and the ferromagnetic semiconductor Cr$_2$Ge$_2$Te$_6$ (CGT)~\cite{Gong-Nature-CGT-2017,Li_CGT-BandStructure2018} encapsulated in hexagonal Boron Nitride (hBN). Low-temperature photoluminescence measurements reveal a partial quenching of the PL and a significantly reduced trion-to-exciton emission ratio in the heterostructure as compared to an isolated MoSe$_2$ ML, indicating an electron transfer from the MoSe$_2$ into the CGT. Under nonresonant, circularly polarized excitation, we find that the exciton-trion ratio depends on the relative orientation of excitation helicity and CGT magnetization. Remarkably, this effect is observable even though the PL emission itself is unpolarized, indicating a vanishing valley polarization of excitons and trions. Density functional theory (DFT) calculations of the heterostructure predict a type-I band alignment and a pronounced spin splitting of the CGT conduction band, yielding a large spin-polarized density of states for interlayer electron tunneling. 
Based on this, we interpret our observations as an interplay of spin-dependent tunneling of the initially valley-polarized electrons from MoSe$_2$ into CGT with exciton and trion formation, valley relaxation and recombination.

Our findings offer a glimpse into material combinations that can be utilized for spin generation and hosting, and more importantly, provide a potential means to read out spin polarization through spin-selective tunneling - an essential component in the process and measurement chain of spin-valleytronic devices.
\section{Results and discussion}

\begin{figure} [h]
    \centering
    \includegraphics[width=1.0\linewidth]{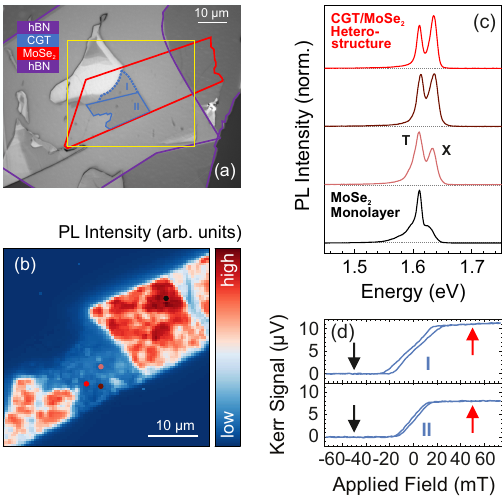}
    \caption{(a) Optical microscope image of an hBN-encapsulated MoSe$_2$/CGT heterostructure. A CGT flake with regions of various layer thickness (I) and (II) covers an underlying MoSe$_2$ monolayer. The yellow box corresponds to the false colour map of the PL scan area (b), showing quenching of the MoSe$_2$ monolayer emission in the heterostructure region. (c) Compared to the MoSe$_2$ monolayer, the ratio of trion (T) and exciton (X) is reduced in the heterostructure, with different exciton-trion ratios observed, depending on the measurement position. All spectra were taken from the PL scan. Their positions are marked by the coloured dots corresponding to the colour of the spectrum. (d) MOKE measurements prove ferromagnetic behaviour of the CGT flake. Magnetic saturation is reached at 32\,mT and 18\,mT for I and II, respectively. The arrows indicate the applied field of $\pm$\,50\,mT for the helicity-resolved measurements described below.}
    \label{Panel1}
\end{figure}

\begin{figure*}
	\centering
	\includegraphics[width=1.0\textwidth]{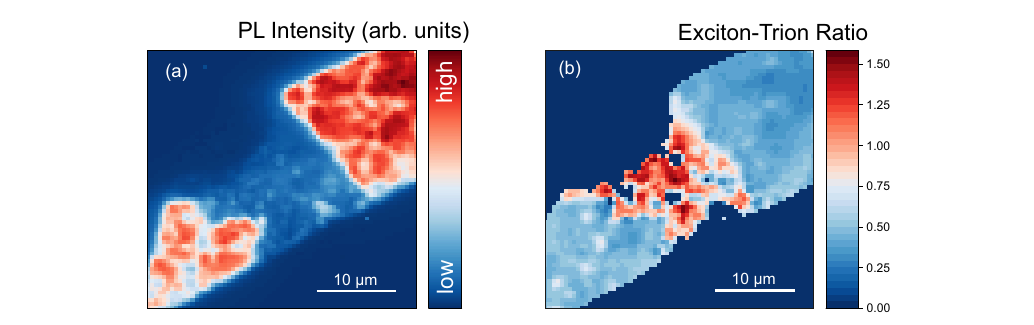}
	\caption{The MoSe$_2$ monolayer emission is quenched in the heterostructure region (a). Here, larger exciton-trion ratios are observed compared to the MoSe$_2$ monolayer (b). The strong scatter between occurring ratios is attributed to inhomogeneous interfacial contact between the MoSe$_2$ monolayer and the CGT flake (see main text).}
	\label{Supplement1}
\end{figure*}

To begin with, we discuss the structure and characterization of our samples. 
As an example, Figure\,\ref{Panel1}a shows a fully hBN-encapsulated MoSe$_2$/CGT heterostructure.
Attached to bulk CGT, two thin CGT regions I and II of different thickness cover an underlying MoSe$_2$ monolayer and are thus of particular interest. After exfoliation, optical images of the CGT flake were taken in transmission mode inside an inert gas glovebox used for sample preparation (see methods). From the absorbance in conjunction with AFM measurements, we estimate the CGT flake to consist of 14 layers in region I and of 9 layers in region II (See Supplementary Section  Figure~\ref{Supplement3}). Low-temperature PL measurements at 5\,K were used to characterize the resulting heterostructure. The PL map of the sample depicted in Figure\,\ref{Panel1}b reveals a quenching by a factor of five of the MoSe$_2$ monolayer emission in the heterostructure region compared to the isolated MoSe$_2$ monolayer, indicating a high-quality interfacial contact for interlayer charge transfer between both layers~\cite{Fang2014}. Whereas the isolated monolayer spectrum is dominated by trion emission, the trion-exciton ratio is significantly smaller in the heterostructure region (Figure\,\ref{Panel1}c). Whether either trion or exciton emission is more pronounced in the spectrum depends on the position on the heterostructure, which we attribute to locally different interface qualities due to inclusions between the two layers. A colour map depicting the spatial distribution of exciton-trion ratios is shown in Figure~\ref{Supplement1}.

For further polarization-resolved PL measurements, a fully saturated magnetization along the out-of-plane axis of the CGT flake is crucial. This makes the magnetization (anti-)parallel to the spin orientation in the TMDC $K$ valleys. 
Polar magneto-optical Kerr effect (MOKE) measurements at a nominal sample temperature of 5\,K (illustrated in Figure\,\ref{Panel1}d) show a hysteresis loop with small remanence and low coercive field for region II. By contrast, we find a more complex, bow-tie-like shape of the hysteresis loop for the thicker region I, which resembles previous reports on hysteresis loops for CGT flakes thicker than 13 layers~\cite{Vervelaki2024_Commun_Mater} or 10\,nm~\cite{Noah2022_NanoLett}, closely matching our thickness estimates based on absorbance.
As a consequence of the small magnetic remanence observed in MOKE, all subsequent measurements were conducted at an external magnetic field of $\pm$\,50\,mT, ensuring magnetization saturation parallel to the direction of the external magnetic field.

\begin{figure*} [t]
    \centering
    \includegraphics[width=\textwidth]{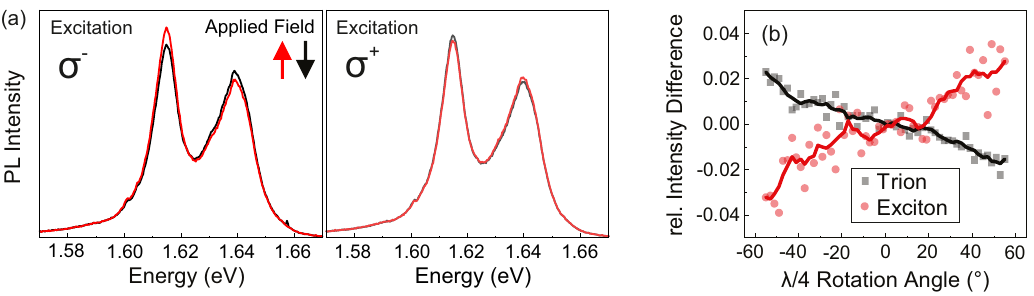}
    \caption{(a) Low-temperature PL spectra taken in the heterostructure region in presence of an external magnetic field ($\pm$\,50\,mT) for left ($\sigma^{-}$) and right ($\sigma^{+}$) circularly polarized excitation. For a fixed excitation helicity, trion and exciton intensities change in dependence of the magnetic field direction (arrows). (b) The relative intensity difference for exciton and trion upon flip of magnetic field depends on the degree of circularly polarized excitation, vanishing for linear polarization at a $\lambda$/4 rotation angle close to 0\,$^\circ$. The solid lines represents the sliding average of five data points and serve as guide to the eye.}
    \label{Panel2}
\end{figure*}

\begin{figure*}
	\centering
	\includegraphics[width=1 \textwidth]{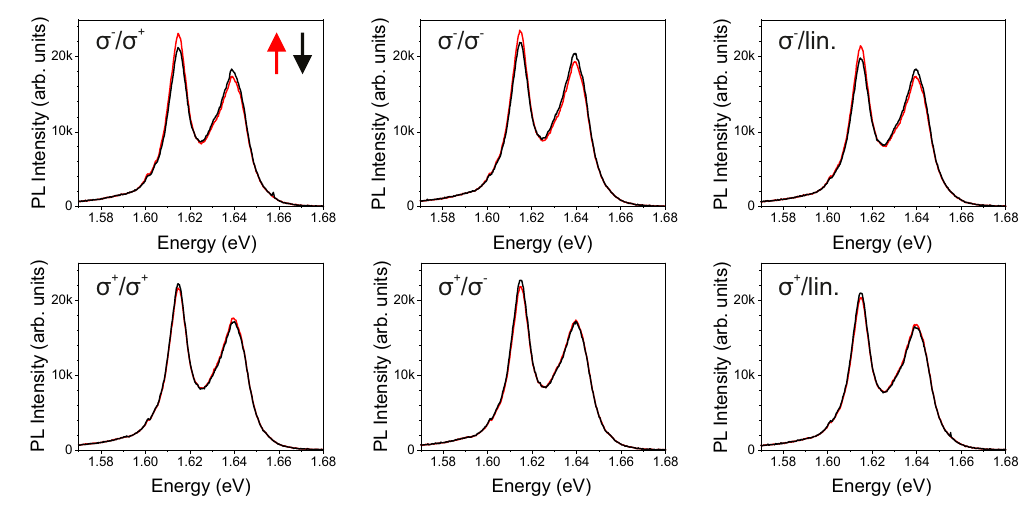}
	\caption{\ PL spectra for different (excitation/detection) helicities and magnetic field directions (arrows). For a constant excitation helicity, the population dynamics is independent of the detection helicity. Measurements were conducted at a magnetic field of $\pm$\,50\,mT.}
	\label{Supplement4}
\end{figure*}

Valley polarization effects were investigated by nonresonant PL measurements using an excitation wavelength of 633\,nm. Circularly polarized excitation was achieved using an achromatic quarter-wave-plate ($\lambda$/4-plate). A second $\lambda$/4-plate in conjunction with a linear polarizer was used for helicity-resolved detection. With a constant excitation helicity, multiple PL spectra were measured with alternating magnetic field direction. Spectra recorded at the same magnetic field direction were then summed up to improve the signal-to-noise ratio. This was repeated for the opposite excitation helicity.
We emphasize that in these measurements the detection helicity was fixed. Remarkably, the population dynamics described below is observed irrespective of the detection helicity and can also be seen in linearly polarized detection (Figure~\ref{Supplement4}). Figure\,\ref{Panel2}a shows evidence for magnetic-field-dependent changes of exciton and trion emission. For left-polarized ($\sigma^{-}$) excitation populating the $K^{-}$ valley, the trion intensity is larger when a positive (red) magnetic field is applied, compared to a negative (black) magnetic field of the same magnitude. The opposite trend is observed for exciton emission where a larger intensity is observed at a negative magnetic field. In contrast, under right-polarized ($\sigma^{+}$) excitation the intensity distributions for both trion and exciton are reversed. However, the magnetic-field-dependent intensity variation has a different magnitude for right and left circularly polarized excitation. We attribute this to a minor deviation from the optimum excitation helicity. In addition, a small beam offset induced by rotating the $\lambda$/4-plate can result in a slightly different excitation spot on the sample. To quantify the normalized magnetic-field-induced intensity differences, we define them as: 
\begin{equation}
    \Delta I = \frac{I(B\downarrow)-I(B\uparrow)}{I(B\downarrow)+I(B\uparrow)}
\end{equation}
where $\downarrow \uparrow$ denotes the external magnetic field direction. Exciton and trion intensities were determined by numerical integration around the respective peak positions. Changing the degree of circular polarization by rotating the $\lambda$/4-plate in the excitation (Figure \ref{Panel2}b), demonstrates a vanishing intensity difference for linear excitation corresponding to an angle of about 0\,$^\circ$, whereas an increasing difference occurs for excitation deviating from linear polarization, reaching maximum values of about 4\,$\%$.

\begin{figure}
	\centering
	\includegraphics[width=0.5\linewidth]{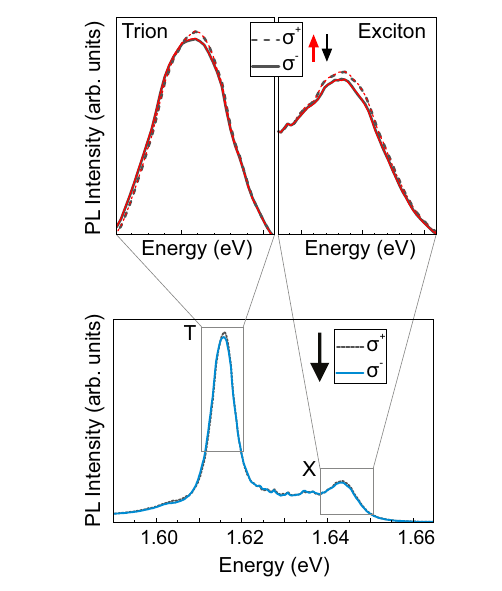}
	\caption{ No magnetic-field-dependent intensity differences for left (solid) and right (dashed line) circularly polarized excitation are detected in the isolated monolayer at a magnetic field of $\pm$\,50\,mT. This proves that the magnetic-field-dependent intensity differences result from an interaction between the MoSe$_2$ monolayer and the overlying CGT layer. For better visibility only the spectra for B$\downarrow$ are shown in the lower panel. The upper panels show a zoom in of trion and exciton peak for both magnetic field directions. Small intensity differences for left and right polarized excitation result from minor changes of laser excitation spot.}
	\label{Supplement2}
\end{figure}

To confirm that the intensity differences result from an interaction between the MoSe$_2$ monolayer and the magnetized CGT, the same measurement was performed on an isolated part of the monolayer. Here, the same trion (exciton) intensity was detected irrespective of the magnetic field direction for both left and right circularly polarized excitation (Figure~\ref{Supplement2}).

We note that we do not observe any discernible Valley Zeeman splitting in our helicity-resolved PL measurements, which were performed using applied magnetic fields of 50~mT, neither for the heterostructure nor for the isolated  MoSe$_2$. This clearly indicates that there is no pronounced proximity-effect-induced Valley Zeeman splitting in our heterostructure, in contrast to previous studies~\cite{Zhong-CrI-WSe-HS2017,Zhang-MoSeCGT-gated2022}.

Based on work functions for both MoSe$_2$ monolayer and CGT flakes of the determined thickness~\cite{Rahman_CGT-WS2021}, the heterostructure should possess a type\,I band alignment with the MoSe$_2$ conduction band being energetically higher than that of CGT, which is verified by DFT calculations. These are discussed in detail in the Supplementary Section (see also references~\cite{ASE,Lazic2015:CPC,Koda2016:JPCC,Carr2020:NRM,Carteaux1995:JP,Schutte1987:JSSC,Hohenberg1964:PRB,Giannozzi2009:JPCM,Kresse1999:PRB,Perdew1996:PRL,Grimme2006:JCC,Grimme2010:JCP,Barone2009:JCC,Zollner2019a:PRB} therein).
PL quenching of the MoSe$_2$ emission in the heterostructure region compared to the isolated monolayer as well as the smaller trion-to-exciton ratio in the heterostructure indicates electron transfer from the MoSe$_2$ into the CGT. We note that there are two mechanisms that can change the exciton-trion ratio in the heterostructure. On one hand, an electron transfer can reduce the background carrier density in MoSe$_2$, reducing the probability of trion formation. On the other hand, trions in MoSe$_2$ have significantly longer photoluminescence lifetimes than excitons~\cite{Wang-valleytimePLMoSe2-2015, Robert-RadLifeTimeTMDs2016}, so that they are more susceptible to nonradiative decay channels, such as electron tunneling into the CGT, during their lifetime. 
\begin{figure} [h]
    \centering
    \includegraphics[width=1.0\linewidth]{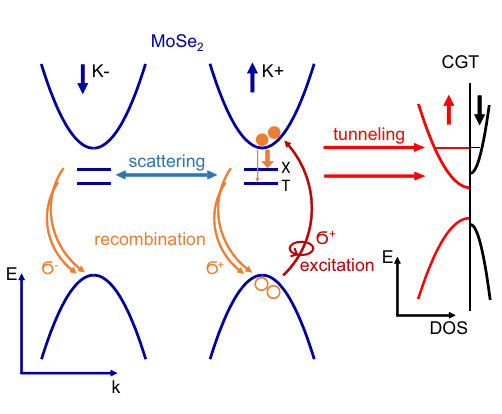}
    \caption{Illustration of processes causing the population dynamics for $\sigma^{+}$ excitation and B\,$\uparrow$. Spin-up electrons in the $K^{+}$ valley tunnel into the CGT due to a high density of spin-up states in the conduction band. The decreased charge carrier density in MoSe$_2$ reduces the trion formation rate compared to excitons. A change of magnetic field direction would result in a smaller density of spin-up states in the CGT, leading to increased trion formation at the expense of excitons. The initial valley polarization is lost due to scattering between the valleys, causing the collected PL to stem from both valleys. The collected PL is thus irrespective of the detection helicity.}
    \label{schematics}
\end{figure}

When analyzing the helicity-dependent changes of exciton and trion emission shown in Figure\,\ref{Panel2}, various processes that occur both within the individual layers and in their interaction have to be considered. We first examine the behaviour for right circularly polarized excitation $(\sigma^{+})$, populating the $K^{+}$ valley, as it is illustrated in Figure\,\ref{schematics}. Due to the external magnetic field, the CGT flake is magnetically saturated with the spin states aligned with the external magnetic field direction. As evident from DFT calculations, a spin-dependent splitting of the CGT conduction band exists (See Figs.~\ref{Fig:layer_resolved_DOS_noSOC} and \ref{Fig:layer_resolved_DOS_bilayerCGT_noSOC} in the Supplementary Section).
In the case of a positive magnetic field, the electrons residing in the $K^{+}$ valley therefore have a high density of spin-up states available in the CGT conduction band. This increases the tunneling rate, which reduces the density of electrons in MoSe$_2$.
Interlayer charge transfer in TMDC heterostructures has been reported, depending on the material combination, to usually occur within 50\,fs~\cite{Hong2014,Jin2018_NatureNanotech}. Similar transfer rates can be expected for TMDC-CGT heterostructures. This is significantly faster than the trion formation time of 2\,ps in MoSe$_2$ monolayers found by Singh et al.~\cite{Singh2016_PhysRevB}. Thus, less electrons are available for trion formation in the MoSe$_2$ monolayer.
In contrast, if a negative magnetic field is applied, the tunneling channel is decreased as a result of the smaller density of states available in the CGT conduction band, and more trions are formed at the expense of excitons. The shifted balance between exciton and trion populations gives rise to an intensity reversal within the spectra upon change of magnetic field direction. Changing the excitation helicity to $\sigma^{-}$ inverts the spin compatibilities with the CGT majority spin states and thus reverses the behaviour explained above. At this point, it shall again be highlighted that the overall detected PL emission is unpolarized. We ascribe this to intervalley scattering mediated by long-range Coulomb interaction~\cite{Glazov2014_Phys_Rev} which was identified to be the main mechanism for exciton valley depolarization after initial circular excitation in MoSe$_2$~\cite{Baranowski-Rates-ValleyPol-2017}. We assume that a considerable fraction of excitons and trions is scattered between the $K^+$ and $K^-$ valleys before exciton and trion recombination take place. 

With fast depolarization rates larger than 1 ps$^{-1}$ for MoSe$_2$ monolayers under weak excitation conditions~\cite{Mahmood2017_NanoLett}, the scattering process competes with or undercuts MoSe$_2$ exciton and trion lifetimes of about 2\,ps and 15\,ps, respectively~\cite{Wang-valleytimePLMoSe2-2015,Robert-RadLifeTimeTMDs2016}. The dependence of the intensity difference on the degree of excitation polarization can directly be explained by the different initial population of the valleys. The imbalance between the two valleys rises as the degree of excitation polarization increases, which is reflected in larger intensity differences upon reversal of the magnetic field. Remarkably, the changes in exciton and trion populations observed in PL are thus a consequence of a spin-dependent tunneling process, even though any initial spin and valley polarization is lost well before exciton and trion recombination.

In summary, we have fabricated MoSe$_2$/CGT heterostructures that reveal charge transfer from the MoSe$_2$ monolayer into the CGT, as predicted from the band alignment resulting from 
DFT calculations.
The magnetization of the CGT flake and resultant spin-split density of states enables spin-dependent tunneling after nonresonant, circularly polarized excitation which manifests itself in altered exciton and trion intensities upon change of the external magnetic field direction. Intervalley scattering causes a loss of initial spin and valley polarization prior to exciton and trion recombination, resulting in an unpolarized collected PL. Our results underline the complex interplay of these competing processes on sub-picosecond timescales.

Given that nonresonant, circularly polarized excitation of our structures yields an exciton/trion occupation imbalance, which is more robust than a valley polarization in MoSe$_2$, our study paves the way for detection of light helicity without the use of polarization optics.

\section{Methods}
\subsection{Sample fabrication}
Under ambient conditions, hBN flakes and MoSe$_2$ monolayers were isolated from bulk crystals by mechanical exfoliation. The hBN flake was stamped with PDMS onto a Si/SiO$_2$ substrate via deterministic transfer~\cite{Castellanos2014}. A MoSe$_2$ monolayer was deposited on the hBN accordingly and afterwards annealed at about 180~$^{\circ}$C in mild vacuum. CGT flakes were exfoliated from a bulk crystal (HQ graphene) under nitrogen atmosphere in a glovebox~\cite{Gant_Sleevebox2020}, thin flakes were thereby identified under an optical microscope based on their optical contrast. Immediately after exfoliation, the CGT flake was placed on top of the beforehand prepared hBN/MoSe$_2$ structure using a second deterministic stamping setup and PDMS transfer inside the glovebox. To protect the CGT from oxidation, the heterostructure was then fully encapsulated inside the glovebox by adding an hBN top layer.

\subsection{Optical measurements}
\subsubsection{Photoluminescence}
For the PL measurements the sample was excited with a 1.96~eV continuous-wave diode laser focused to a spot size of about 1\,µm using an 80x microscope objective. The sample was mounted in a He-flow cryostat and cooled to a nominal temperature of about 5\,K. To prevent sample heating, the excitation density was kept below 4\,kW/cm$^2$.
The PL light emitted by the sample was collected using the same objective, filtered by long pass and analyzed with a combination of a spectrometer and a charge-coupled-device. 
To obtain PL maps of the sample the cryostat, with the sample inside, was moved in relation to the fixed laser spot through a computer-controlled xy stage. For helicity-resolved excitation, a linear polarizer and an achromatic quarter-wave plate ($\lambda$/4) were placed in the excitation beam path. The wave plate was mounted in a motorized rotation stage, so that its angle could be varied automatically. Similarily, for helicity-resolved detection, an achromatic $\lambda$/4-plate and a linear polarizer (acting as an analyzer) were placed in the detection beam path in front of the spectrometer. 
For application of external magnetic fields, an air coil was placed around the cryostat, so that magnetic fields of up to 200~mT could be applied perpendicular to the sample plane. The current for the coil was supplied by a bipolar current source. 
\subsubsection{Polar magneto-optical Kerr effect}
The polar MOKE measurements were performed in the setup described above using the 1.96~eV diode laser applying an excitation density of about  1\,kW/cm$^2$. All measurements shown in the manuscript were performed at a nominal temperature of about 5\,K. Laser intensity was modulated using a flywheel chopper. A beamsplitter cube was introduced into the detection beam path, so that the reflected laser light could be guided to an optical bridge detector with a pair of balanced photodiodes (see~\cite{Michi-Kerr2021} for details). The difference signal from the photodiodes was detected using a lock-in amplifier. MOKE loops were measured by tracking the difference signal as a function of the applied magnetic field controlled by the bipolar current source. 
\\
\section{Acknowledgements}
The authors gratefully acknowledge technical assistance by E. Moldt and fruitful discussions with R. Edhib.
S. D. acknowledges financial support by the Humboldt foundation and a startup funding grant provided by the Deutsche Forschungsgemeinschaft (DFG, German Research Foundation) \emph{via} SPP2244. T.K. acknowledges financial support by the DFG \emph{via} the following grants: SFB1477 (project No. 441234705), SPP2244 (project No. 443361515),  KO3612/7-1 (project No. 467549803) and KO3612/8-1 (project No. 549364913). K. Z. and J. F. acknowledge funding by the DFG \emph{via}  SFB 1277 (Project No. 314695032) and SPP 2244 (Project No. 443416183), as well as the European Union Horizon 2020 Research and Innovation Program under contract number 881603 (Graphene Flagship) and FLAGERA project 2DSOTECH. K.W. and T.T. acknowledge support from the JSPS
KAKENHI (grant numbers 21H05233 and 23H02052) and World Premier
International Research Center Initiative (WPI), MEXT, Japan.

\onecolumngrid
\newpage
\section{Supplementary experimental data}
\begin{figure*}[b]
	\centering
	\includegraphics[width=0.7\textwidth]{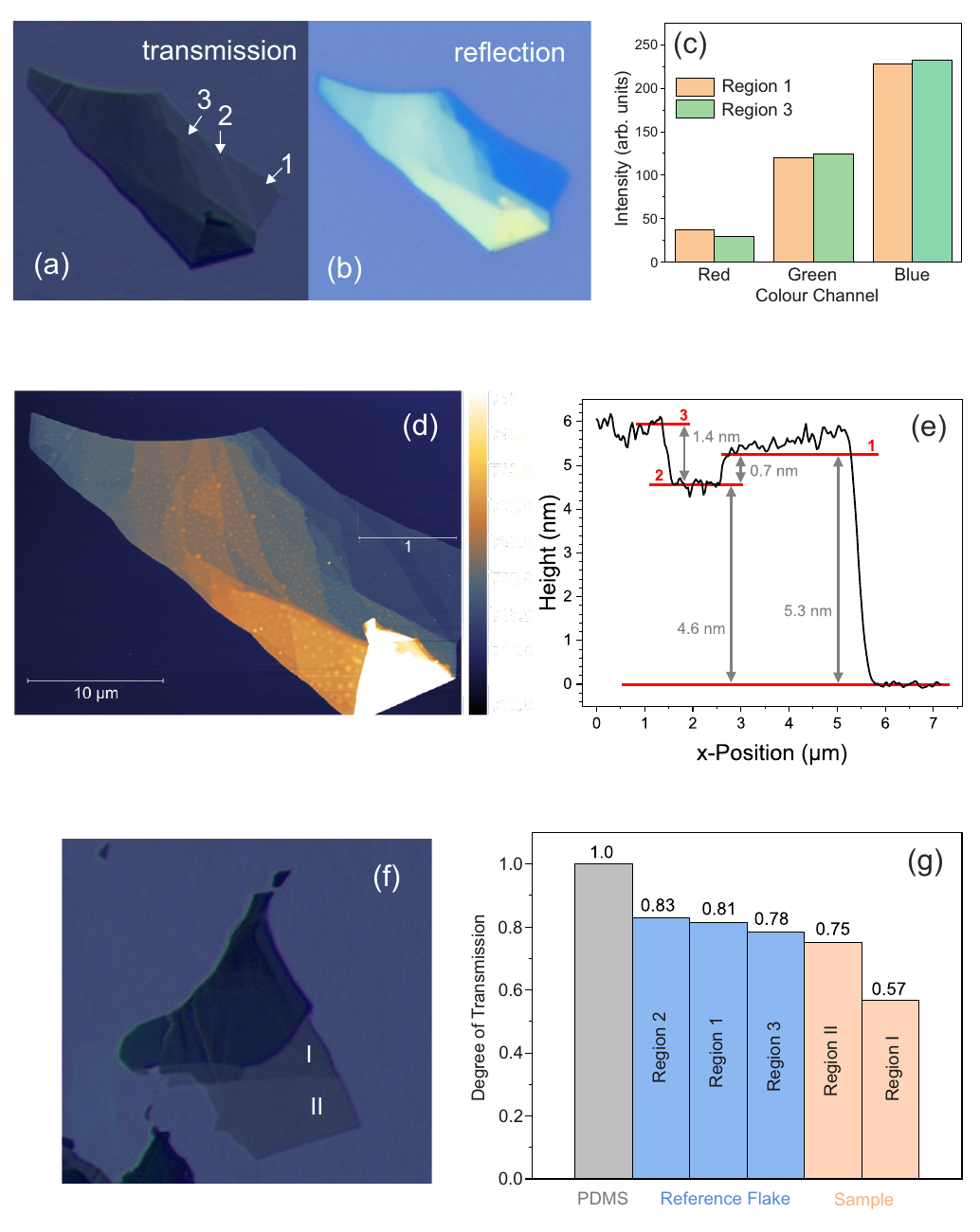}
	\caption{ The CGT sample thickness was determined using a reference sample. (a) The transmission microscopy image of the reference sample on PDMS shows three thin regions (1-3) of different thickness. Here, a clear contrast between 1 and 2 is visible, whereas the contrast between 1 and 3 is poor. The reflection light image (b) of the same sample on a Si/SiO$_2$ substrate reveals a difference in layer thickness between 1 and 3 which manifests itself in different nuances of blue. The contrast is reflected in all colour channels (c). (d,e) AFM measurements were performed to elucidate the actual sample thickness. Assuming a monolayer step height of 1.1\,nm and a step height of 0.7\,nm for every subsequent layer in accordance with~\cite{Gong-Nature-CGT-2017}, region 2 consists of 6, region 1 of 7 and region 3 of 8 layers. (f,g) From the transmission microscopy images the degree of transmission for the reference flake and the CGT flake shown in the main text was determined, using the PDMS film as a calibration. We assume a reduction of about 3 percent per layer. A comparison between the values for the flake shown in the main text and the reference flake hints at layer thicknesses of 9 layers in region I and 14 layers in region II.}
	\label{Supplement3}
\end{figure*}

\newpage

\clearpage
\newpage
\newpage
\section{DFT Calculations on MoSe$_2$/Cr$_2$Ge$_2$Te$_6$}

\subsection{Structural Setup}
\begin{figure}[!htb]
	\includegraphics[width=0.99\columnwidth]{Structure.pdf}
	\caption{Top and side view of the MoSe$_2$/Cr$_2$Ge$_2$Te$_6$ heterostructure. The supercell has 69 atoms, with the lattice parameters of $|a| = |b| = 11.855$~\AA~and $|c| = 34.065$~\AA. The relaxed average interlayer distance between the layers is $d = 3.515$~\AA. The purple (blue) shaded area indicates the monolayer MoSe$_2$ (Cr$_2$Ge$_2$Te$_6$) unit cell, while the black dashed line is the heterostructure unit cell. The twist angle between the layers is 16.102°. 
		\label{Fig:Structure}}
\end{figure}
The MoSe$_2$/Cr$_2$Ge$_2$Te$_6$ heterostructure was set-up with the {\tt atomic simulation environment (ASE)} \cite{ASE} and the {\tt CellMatch} code \cite{Lazic2015:CPC}, implementing the coincidence lattice method \cite{Koda2016:JPCC,Carr2020:NRM}. The lattice constant of Cr$_2$Ge$_2$Te$_6$ within the heterostructure is 6.8445~\AA, slightly strained by 0.25\%~\cite{Carteaux1995:JP}. The MoSe$_2$ layer is kept unstrained with a lattice constant of 3.288~\AA~\cite{Schutte1987:JSSC}. Therefore, the individual monolayers are barely strained in our heterostructure and we should be able to reliably extract band offsets as well as proximity exchange effets. In order to simulate quasi-2D systems, we add a vacuum of about $18$~\AA~to avoid interactions between periodic images in our slab geometry. The resulting heterostructure is shown in Fig.~\ref{Fig:Structure}.

\subsection{Computational Details}

The electronic structure calculations and structural relaxations of the MoSe$_2$/Cr$_2$Ge$_2$Te$_6$ heterostructure is performed by DFT~\cite{Hohenberg1964:PRB} 
with {\tt Quantum ESPRESSO}~\cite{Giannozzi2009:JPCM}. Self-consistent calculations are carried out with a $k$-point sampling of $24\times 24\times 1$. 
We perform open shell calculations that provide the
spin-polarized ground state of the Cr$_2$Ge$_2$Te$_6$ monolayer. 
A Hubbard parameter of $U = 1.0$~eV is used for Cr $d$-orbitals, being in the range of proposed $U$ values especially for this compound \cite{Gong-Nature-CGT-2017}.
We use an energy cutoff for charge density of $600$~Ry and the kinetic energy cutoff for wavefunctions is $75$~Ry for the (scalar) relativistic pseudopotentials
with the projector augmented wave method~\cite{Kresse1999:PRB} with the 
Perdew-Burke-Ernzerhof exchange correlation functional~\cite{Perdew1996:PRL}.
For the relaxation of the heterostructures, we add DFT-D2 vdW corrections~\cite{Grimme2006:JCC,Grimme2010:JCP,Barone2009:JCC} and use 
quasi-Newton algorithm based on trust radius procedure. 
To get proper interlayer distances and to capture possible moir\'{e} reconstructions, we allow all atoms to move freely within the heterostructure geometry during relaxation. Relaxation is performed until every component of each force is reduced below $5\times10^{-4}$~[Ry/$a_0$], where $a_0$ is the Bohr radius.

\subsection{Results}

\subsubsection{Without Spin-Orbit Coupling}
\begin{figure}[!htb]
	\includegraphics[width=0.99\columnwidth]{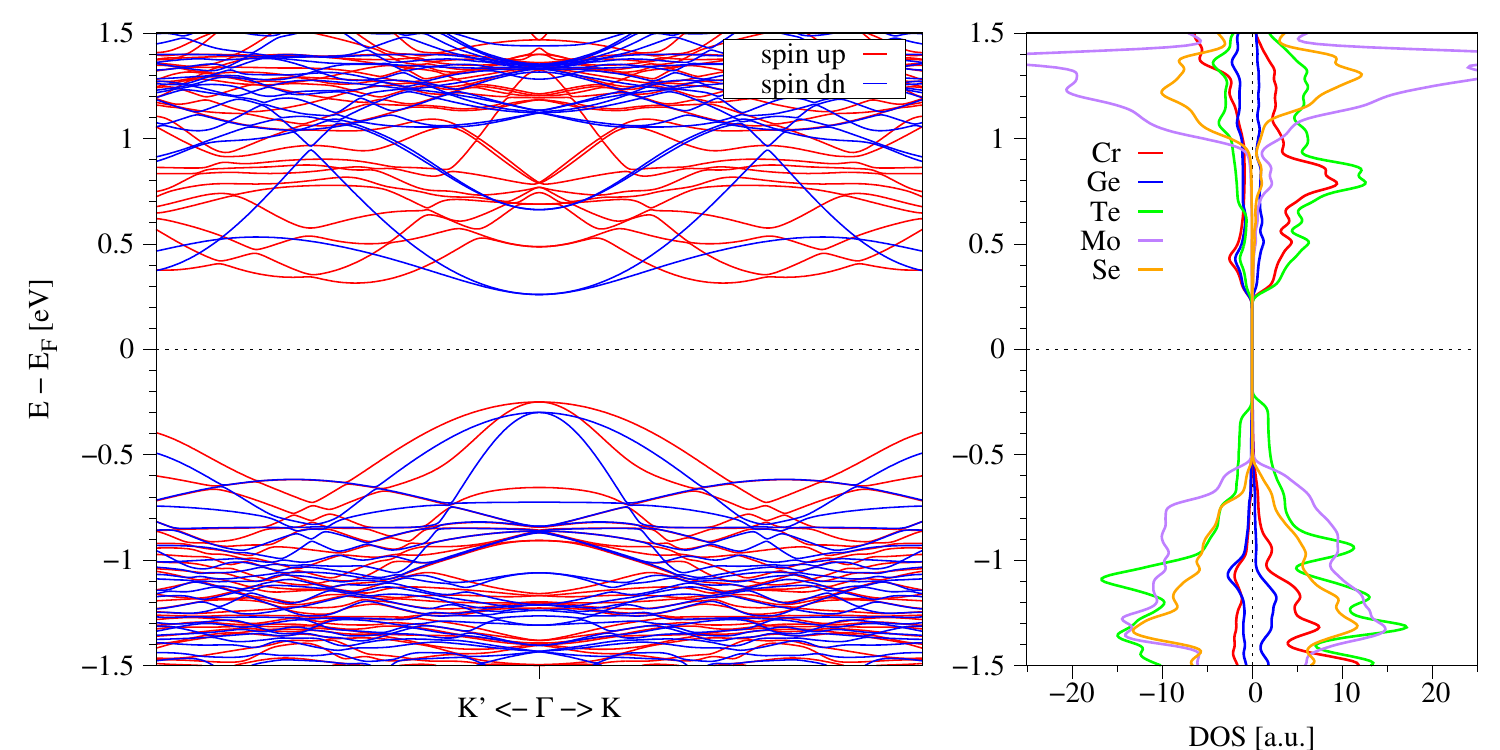}
	\caption{Left: DFT-calculated band structure of the MoSe$_2$/Cr$_2$Ge$_2$Te$_6$ heterostructure towards the MoSe$_2$ valley edges at K/K'.  Red (blue) lines correspond to spin up (down).
		Right: The corresponding spin and atom resolved density of states. Positive (negative) DOS is for spin up (down).
		\label{Fig:bands_dos_noSOC}}
\end{figure}

First, we discuss the first-principles results, where spin-orbit coupling is excluded from the calculations and the magnetization direction of 
Cr$_2$Ge$_2$Te$_6$ is fixed to be collinear with the $z$-axis (transverse to the interface).
In Fig.~\ref{Fig:bands_dos_noSOC} we show the calculated band structure and density of states (DOS) of the MoSe$_2$/Cr$_2$Ge$_2$Te$_6$ heterostructure. We find that the heterostructure forms a type I band alignment, as the Cr$_2$Ge$_2$Te$_6$ band edges reside within the MoSe$_2$ band edges. The calculated dipole of the heterostructure is -0.0542 debye, so there is almost no electric field across the interface.
The averaged induced magnetic moments on the different atomic layers are: Cr $= 3.484~\mu_{\textrm{B}}$, Ge $= 0.058~\mu_{\textrm{B}}$, Te $= -0.192~\mu_{\textrm{B}}$, Se$_{1}$ $= -0.0015~\mu_{\textrm{B}}$, Mo $= -0.002~\mu_{\textrm{B}}$, and Se$_{2}$ $= -0.0001~\mu_{\textrm{B}}$.

\begin{figure}[!htb]
	\includegraphics[width=0.99\columnwidth]{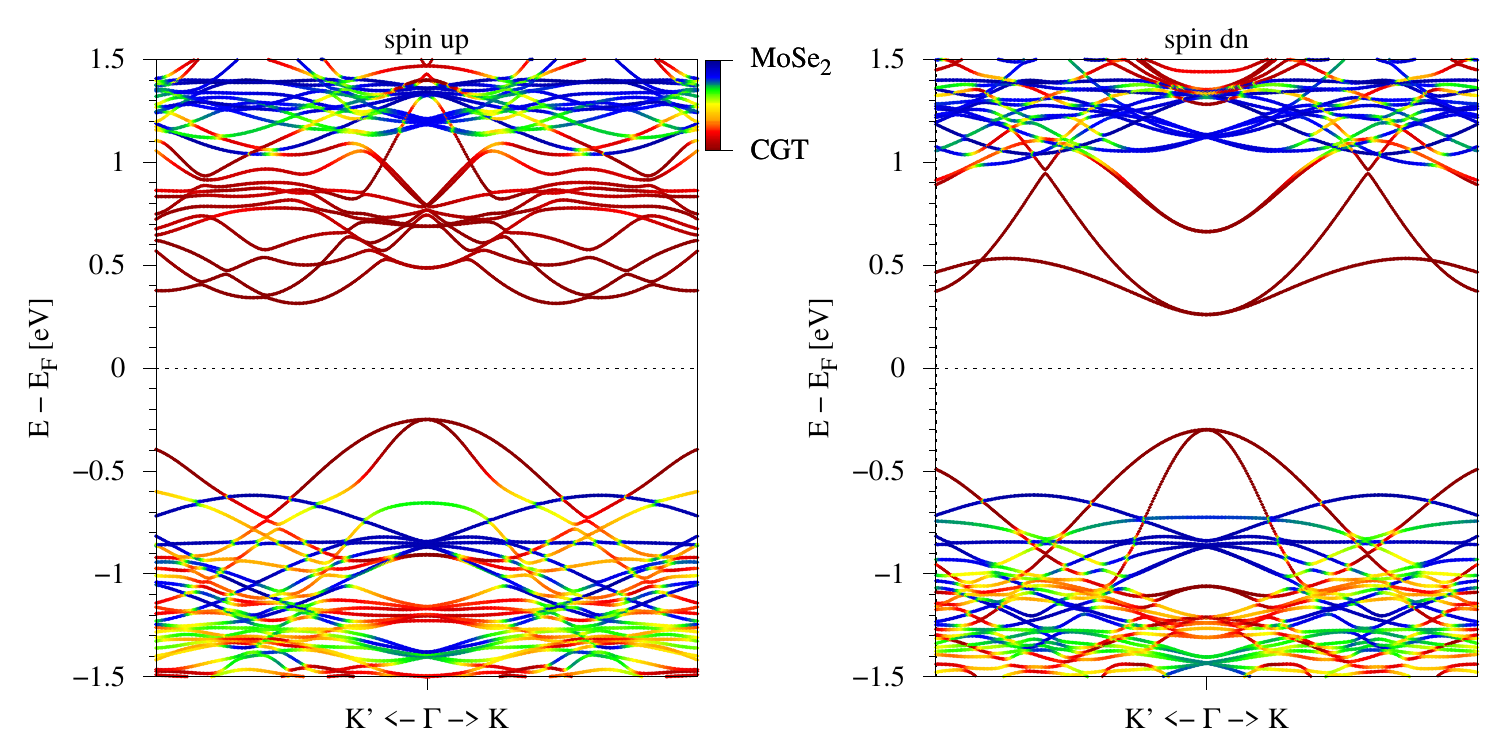}
	\caption{The layer-projected band structure for spin up (left) and spin down (right) channels. The color code indicates the projection onto the monolayers. 
		\label{Fig:bands_projection_noSOC}}
\end{figure}

\begin{figure}[!htb]
	\includegraphics[width=0.99\columnwidth]{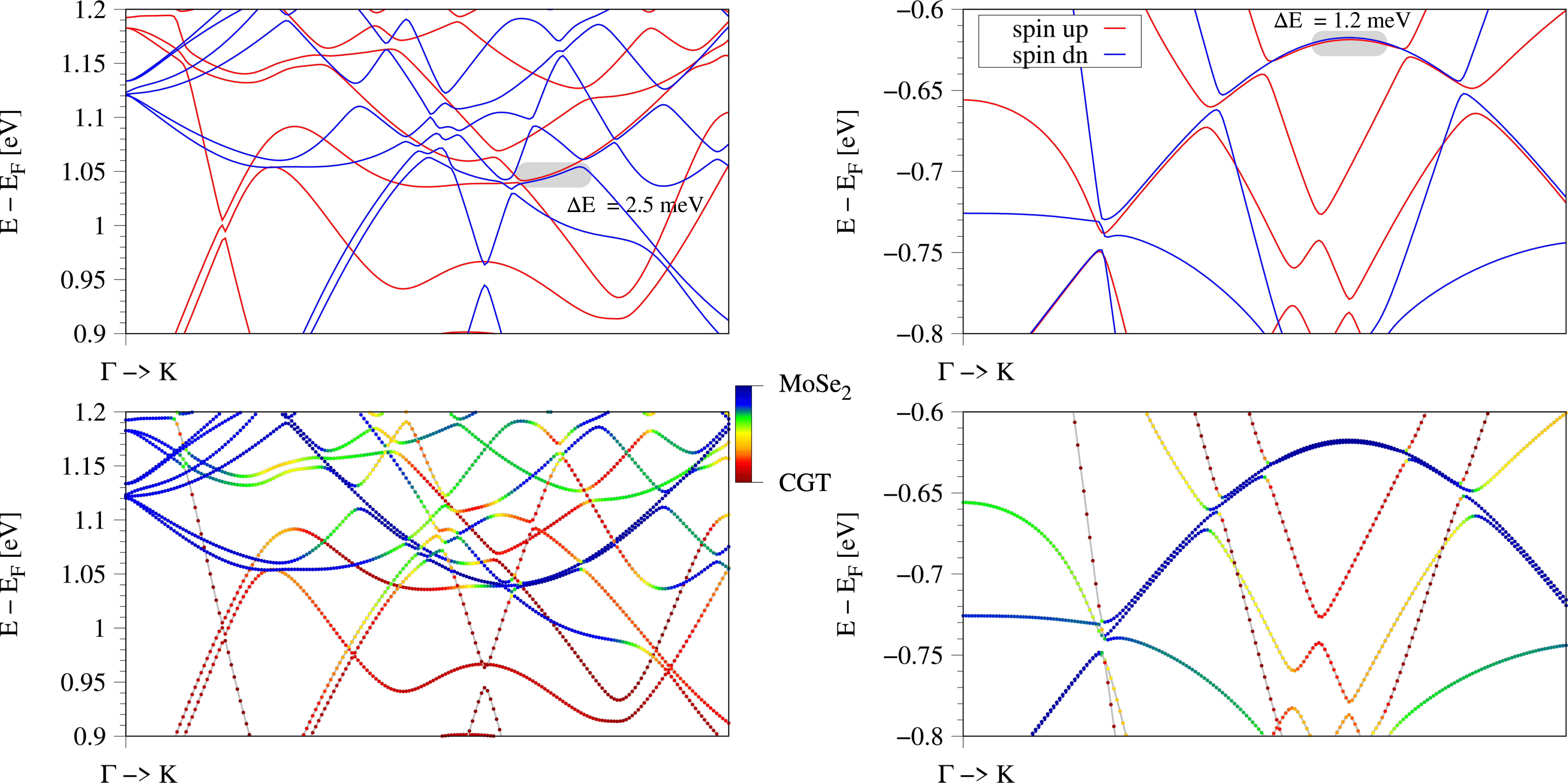}
	\caption{Zoom to the band edges near the K-valley of MoSe$_2$.  
		\label{Fig:bands_zoom_noSOC}}
\end{figure}

In particular the DOS gives a first indication of a strong hybridization of conduction band states. 
To further confirm this, we have calculated a layer-projected band structure, shown in Fig.~\ref{Fig:bands_projection_noSOC}.
The valence band edges of MoSe$_2$ are almost unperturbed within the heterostructure, but the conduction band edge states are rather strongly hybridized. A detailed zoom to the MoSe$_2$ band edges is given in Fig.~\ref{Fig:bands_zoom_noSOC}. Due to proximity-induced exchange coupling, the band edges experience spin splittings, as indicated. Since the valence band edge of MoSe$_2$ rather weakly hybridizes (anticrosses) with Cr$_2$Ge$_2$Te$_6$ bands, the splitting is about 1.2~meV. In contrast, the conduction band edge of MoSe$_2$ is much stronger hybridized leading to a spin-splitting of about 2.5~meV near the band edge. 

\begin{figure}[!htb]
	\includegraphics[width=0.99\columnwidth]{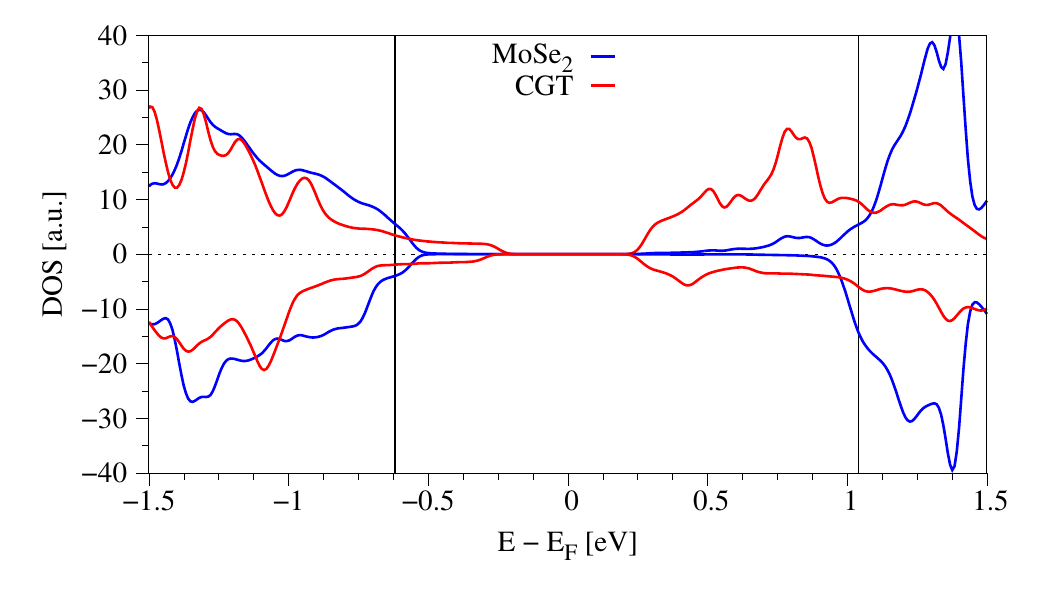}
	\caption{The calculated layer-resolved density of states of the MoSe$_2$/Cr$_2$Ge$_2$Te$_6$ heterostructure. Positive (negative) DOS is for spin up (down). Black vertical lines indicate the MoSe$_2$ band edges at K/K'.
		\label{Fig:layer_resolved_DOS_noSOC}}
\end{figure}

Experimentally, the hybridization opens an efficient charge transfer channel for photo-excited charge carriers from MoSe$_2$ into Cr$_2$Ge$_2$Te$_6$. 
Since most of the heterostructure bands below the MoSe$_2$ conduction band edge are spin-up polarized Cr$_2$Ge$_2$Te$_6$ bands, a predominant transfer of spin-up polarized electrons is expected. The predominant tunneling of spin-up electrons is supported by the layer-resolved density of states, see Fig.~\ref{Fig:layer_resolved_DOS_noSOC}.

\begin{figure}[!htb]
	\includegraphics[width=0.99\columnwidth]{CDD.pdf}
	\caption{DFT-calculated charge redistribution, by subtracting the monolayer charge densities from the heterostructure one. The colors yellow and gray correspond to gain and loss of charge. The isovalue is set to $1\times10^{-4}$ e/\AA$^3$. 
		\label{Fig:CDD}}
\end{figure}

\clearpage
\subsubsection{With Spin-Orbit Coupling, Noncollinear Case}

To get a more realistic description of the heterostructure dispersion, we consider noncollinear magnetism and spin-orbit coupling in the calculations. The reason is that spin-orbit coupling is utterly important for the description of MoSe$_2$. 
Additionally, the noncollinear calculation allows the atoms to freely adjust their magnetization directions, as interfacial hybridization may disturb the out-of-plane magnetism of Cr$_2$Ge$_2$Te$_6$.

\begin{figure}[!htb]
	\includegraphics[width=0.99\columnwidth]{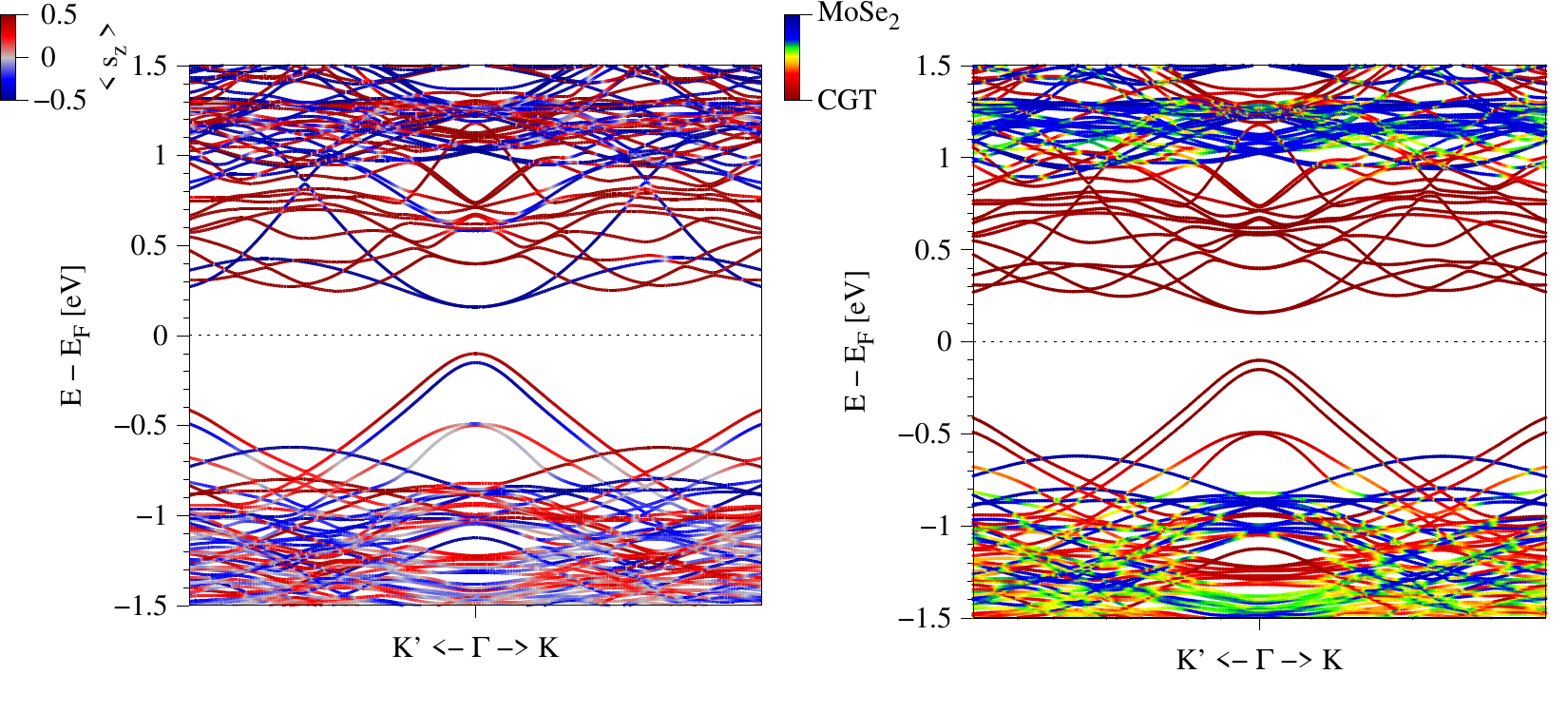}
	\caption{DFT-calculated band structure of the MoSe$_2$/Cr$_2$Ge$_2$Te$_6$ heterostructure, with spin-orbit coupling and including noncollinear magnetism. We show the bands with $s_{z}$ and layer projections.
		\label{Fig:bands_spins_proj_SOC}}
\end{figure}

\begin{figure}[!htb]
	\includegraphics[width=0.99\columnwidth]{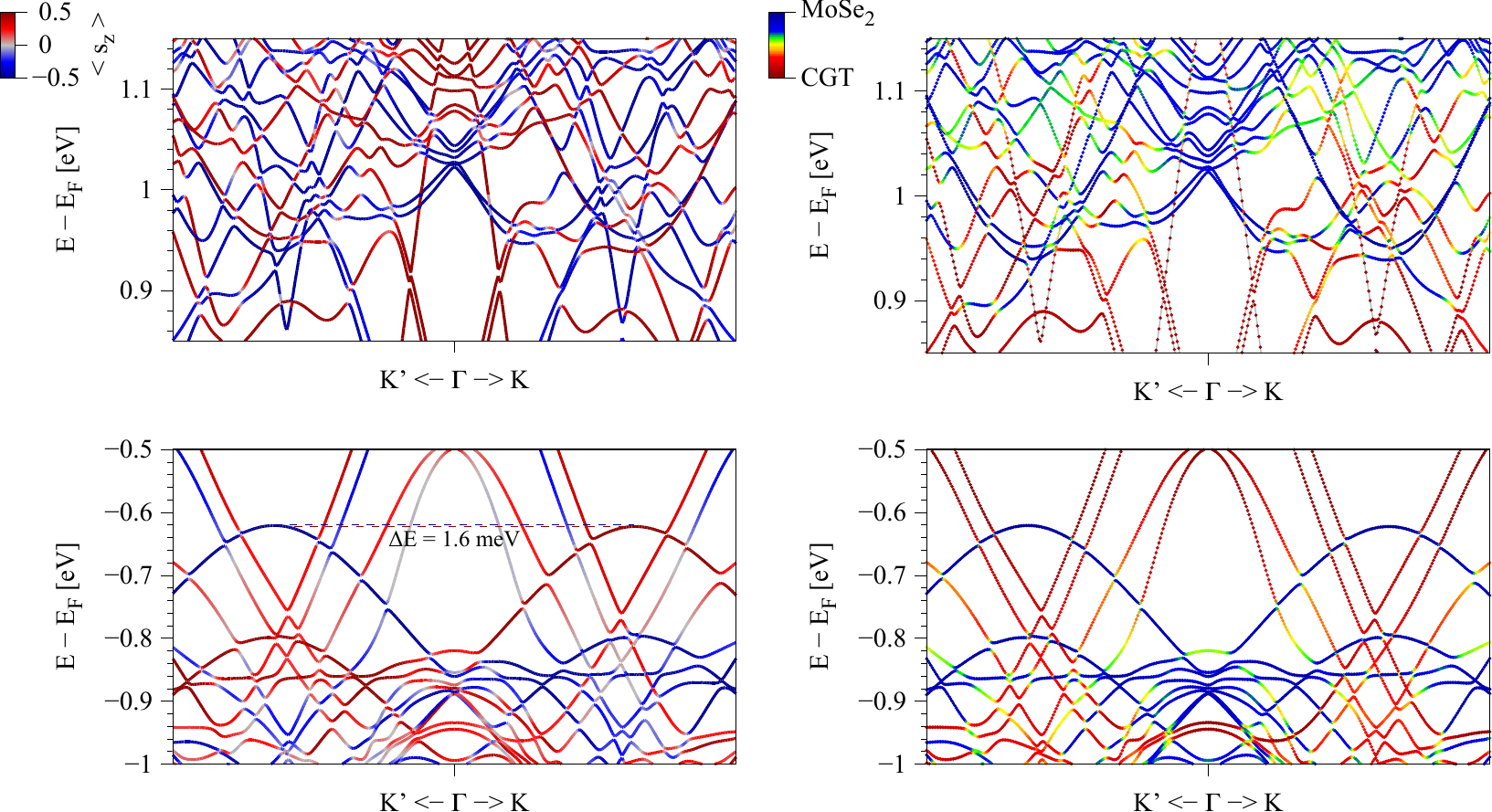}
	\caption{Zoom to the band edges near K/K' valleys of MoSe$_2$.
		\label{Fig:bands_spins_proj_zoom_SOC}}
\end{figure}

In Fig.~\ref{Fig:bands_spins_proj_SOC}, we show the band structure of the heterostructure including spin-orbit coupling. The type-I band alignment remains, as we find from the layer-projected dispersion. 
The spin character of the bands is mainly of $s_z$, indicating that the MoSe$_2$ Ising spin-orbit coupling and the out-of-plane magnetism of Cr$_2$Ge$_2$Te$_6$ also remain intact. 
Zooming into the band edges of MoSe$_2$, see Fig.~\ref{Fig:bands_spins_proj_zoom_SOC}, we find that the valence band edges at K/K' are well preserved, but are energetically split by about 1.6~meV, stemming from the proximity-induced magnetism. In contrast, the conduction band edges can  also still be recognized, but are well hybridized. 
Apart from the additional spin-orbit coupling induced band splittings, the overall picture remains the same, supporting spin-dependent tunneling of photo-excited electrons from MoSe$_2$ into Cr$_2$Ge$_2$Te$_6$. 

\clearpage
\subsubsection{Increasing the thickness of the magnet}
\begin{figure}[!htb]
	\includegraphics[width=0.99\columnwidth]{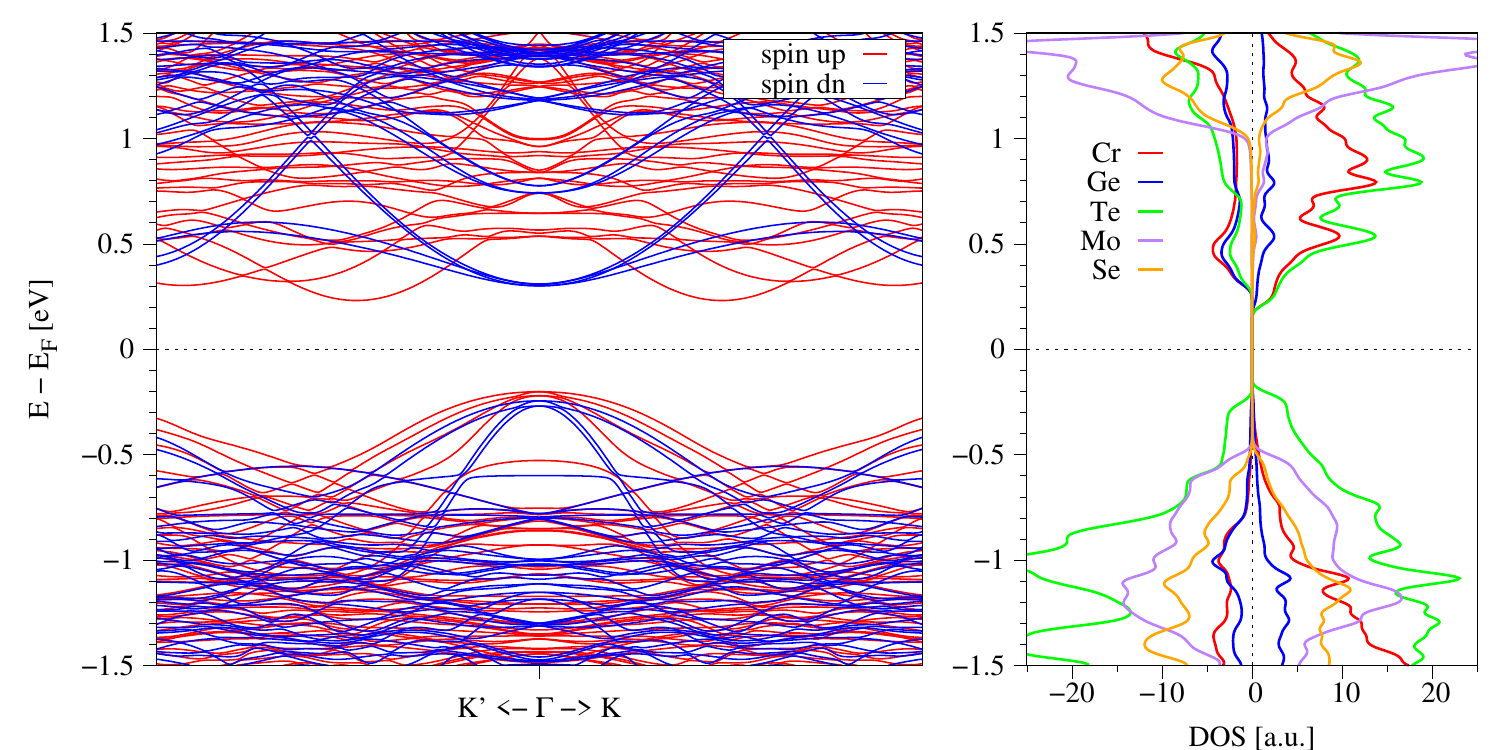}
	\caption{Left: DFT-calculated band structure of the MoSe$_2$/bilayer-Cr$_2$Ge$_2$Te$_6$ heterostructure towards the MoSe$_2$ valley edges at K/K'.  Red (blue) lines correspond to spin up (down).
		Right: The corresponding spin and atom resolved density of states. Positive (negative) DOS is for spin up (down).
		\label{Fig:bands_dos_noSOC_bilayerCGT}}
\end{figure}

\begin{figure}[!htb]
	\includegraphics[width=0.99\columnwidth]{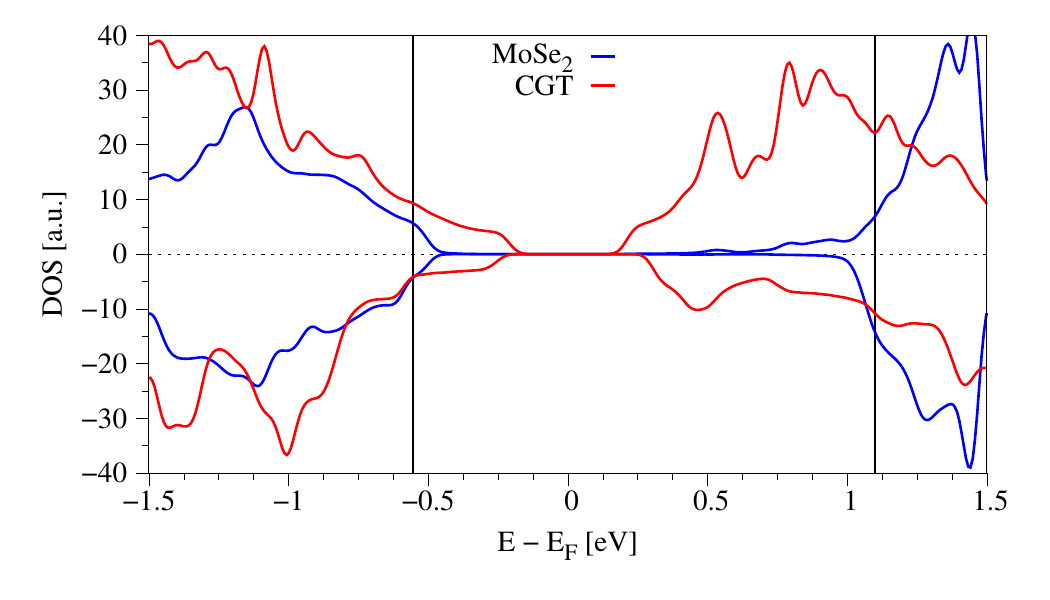}
	\caption{The calculated layer-resolved density of states of the MoSe$_2$/bilayer-Cr$_2$Ge$_2$Te$_6$ heterostructure. Positive (negative) DOS is for spin up (down). Black vertical lines indicate the MoSe$_2$ band edges at K/K'.
		\label{Fig:layer_resolved_DOS_bilayerCGT_noSOC}}
\end{figure}

\begin{figure}[!htb]
	\includegraphics[width=0.99\columnwidth]{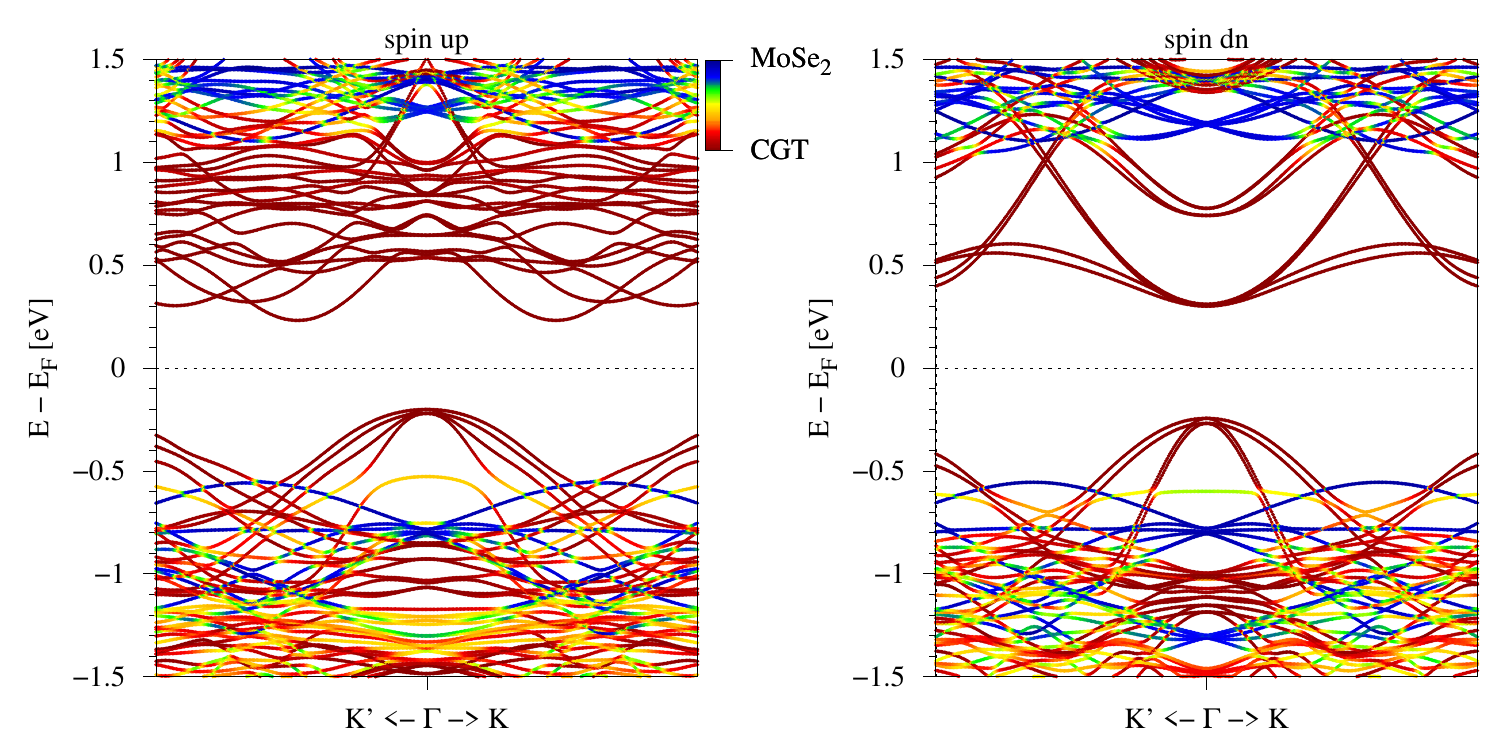}
	\caption{The layer-projected band structure for spin up (left) and spin down (right) channels. The color code indicates the projection onto the monolayers. 
		\label{Fig:bands_projection_noSOC_bilayerCGT}}
\end{figure}

\begin{figure}[!htb]
	\includegraphics[width=0.99\columnwidth]{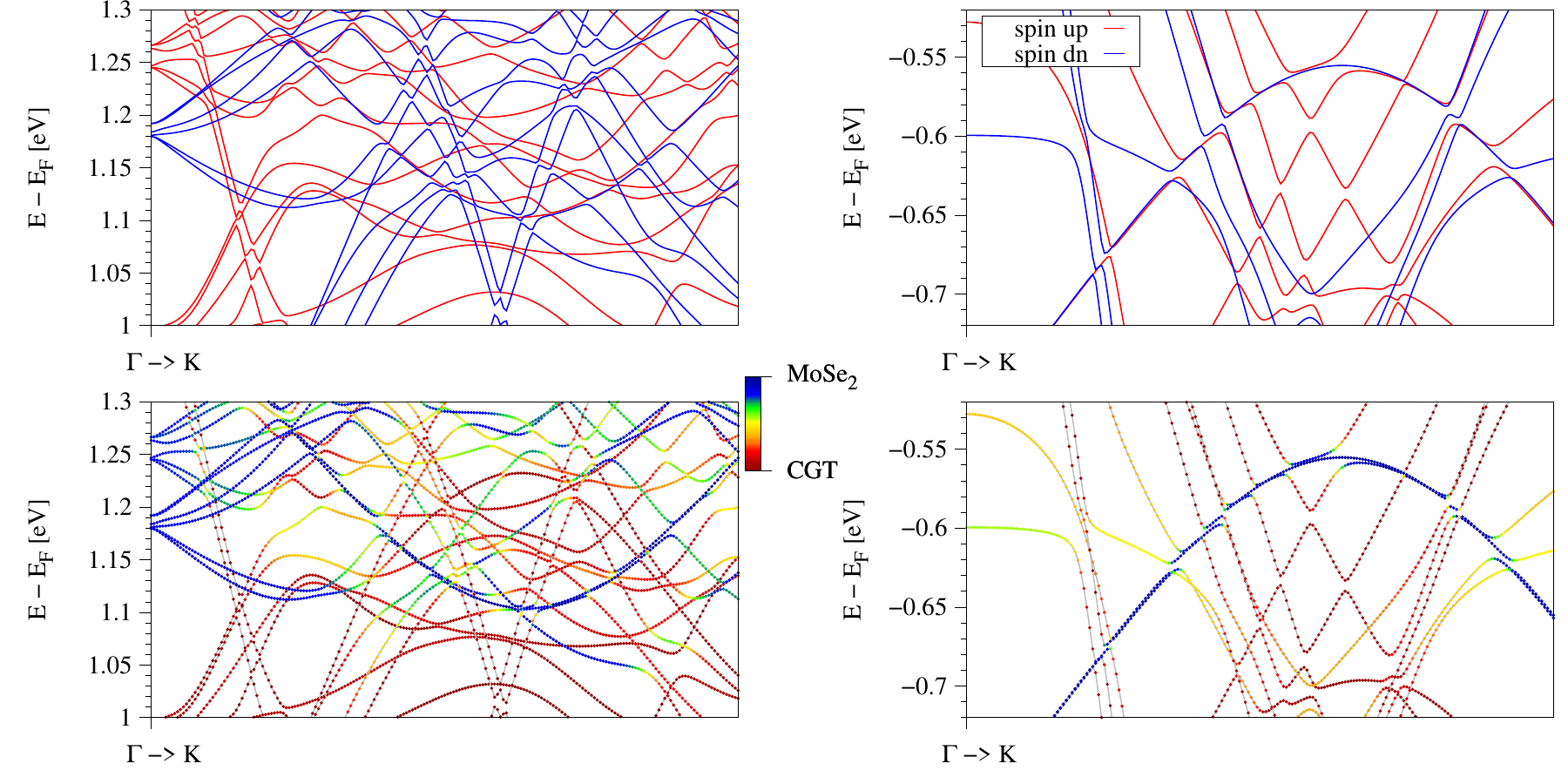}
	\caption{Zoom to the band edges near the K-valley of MoSe$_2$.  
		\label{Fig:bands_zoom_noSOC_bilayerCGT}}
\end{figure}

Experimentally, MoSe$_2$ is considered on few (9-15) layers of Cr$_2$Ge$_2$Te$_6$. 
Initial experiments on the thickness-dependence of the Curie temperature of Cr$_2$Ge$_2$Te$_6$, indicate that this thickness can be regarded as bulk crystal ~\cite{Gong-Nature-CGT-2017}. Computationally it is very demanding to simulate heterostructures, since many atoms are involved in the geometry. However, we know from previous studies that the proximity-coupling is short-ranged and typically restricted to neighboring layers~\cite{Zollner2019a:PRB}. Therefore, we add a second Cr$_2$Ge$_2$Te$_6$ layer beneath the first one, in AB-stacking configuration and without changing the interfacial alignment of the MoSe$_2$/Cr$_2$Ge$_2$Te$_6$ heterostructure, and look for qualitative changes in the results. Before calculating the dispersion, we allow for full structural relaxation of this trilayer system. 
Based on the results in Figs. \ref{Fig:bands_dos_noSOC_bilayerCGT}, \ref{Fig:layer_resolved_DOS_bilayerCGT_noSOC}, \ref{Fig:bands_projection_noSOC_bilayerCGT}, and \ref{Fig:bands_zoom_noSOC_bilayerCGT}, we find barely any qualitative difference to the monolayer case, indicating the short-rangeness of proximity coupling.

\end{document}